**Mirishli Shahmar Sakit,**
LL.M and Lawyer
Address: A.Naxchivani 13C 96, AZ1125, Baku
E-mail: Shahmarmirishli@aidatachronicles.com

# REGULATING AI IN FINANCIAL SERVICES: LEGAL FRAMEWORKS AND COMPLIANCE CHALLENGES

***Açar sözlər:*** *süni intellektin tənzimlənməsi, maliyyə xidmətlərinə uyğunluq, alqoritmik idarəetmə, məlumatların qorunması qanunu, Sİ etikası, tənzimləyici texnologiya, fintex qanunu.*

***Ключевые слова:*** *регулирование искусственного интеллекта, соответствие требованиям финансовых услуг, алгоритмическое управление, закон о защите данных, этика ИИ, регуляторные технологии, финтех-право.*

The financial services industry is undergoing a profound transformation driven by the rapid advancement and integration of Artificial Intelligence (AI) technologies. From algorithmic trading and fraud detection to personalized financial advice and credit scoring, AI is fundamentally reshaping how financial institutions operate, make decisions, and interact with their clients [14, p. 6]. The allure of increased efficiency, reduced costs, and enhanced decision-making capabilities has spurred widespread adoption of AI across the financial landscape, promising to unlock new levels of performance and customer service [9, p. 1].

AI technologies, including machine learning, natural language processing, and deep learning, are being deployed across a wide spectrum of financial activities. These applications range from high-frequency trading algorithms that can execute thousands of trades per second to sophisticated credit scoring models that analyze vast arrays of alternative data to assess creditworthiness [3, p. 2]. AI-powered chatbots and virtual assistants are revolutionizing customer service, while advanced analytics are enhancing risk management and regulatory compliance processes [8, p. 15].

The potential benefits of AI in finance are substantial. By leveraging vast amounts of data and complex algorithms, AI systems can identify patterns, predict market movements, and make decisions with a speed and accuracy that surpasses human capabilities. This enhanced analytical power promises to improve market efficiency, expand access to financial services for underserved populations, and strengthen the overall stability of the financial system [15, p. 731].

However, the integration of AI in finance also introduces significant risks and challenges that demand careful scrutiny. One of the primary concerns is the potential for algorithmic bias and discrimination. AI systems, trained on historical data, may inadvertently perpetuate or even amplify existing biases, leading to unfair and discriminatory outcomes in credit decisions, insurance pricing, and other critical financial services [22, p. 132]. The potential for AI to exacerbate financial exclusion and inequality poses a serious ethical and legal challenge that regulators must address.

The vast amounts of personal and financial data required to train and operate AI systems raise critical concerns about data protection, privacy rights, and the potential for data breaches or misuse [20, p. 8]. As AI systems become more sophisticated in their ability to analyze and infer sensitive information from seemingly innocuous data points, the boundaries of privacy protection are being pushed to their limits.

The complexity and opacity of many AI algorithms, often referred to as the "black box" problem, pose significant challenges for accountability and regulatory oversight [15, p. 3]. The inability to fully explain or interpret AI-driven decisions undermines principles of fairness and due process, particularly in high-stakes financial decisions that significantly impact individuals' lives.

As financial institutions increasingly rely on AI for critical functions such as risk assessment and trading, there is a growing concern about the potential for AI systems to introduce new forms of systemic risk [9, p. 2]. The interconnectedness of AI-driven financial systems and the potential for cascading errors or unexpected interactions between algorithms raise important questions about financial stability and market integrity.

The rapid pace of AI innovation in finance is outstripping the ability of regulatory frameworks to keep up, creating potential gaps in oversight and consumer protection [5, p. 3]. Regulators face the daunting task of developing frameworks that can effectively govern AI technologies while remaining flexible enough to accommodate future innovations.

Regulatory bodies around the world are grappling with how to strike the right balance between fostering innovation and protecting consumers and market integrity. This has led to a diverse landscape of regulatory approaches, from comprehensive AI-specific legislation to more sector-specific guidelines and principles.

In the European Union, the Artificial Intelligence Act, which came into effect in August 2024, represents the most ambitious attempt to create a comprehensive regulatory framework for AI [10]. The Act adopts a risk-based approach, categorizing AI systems based on their level of risk, with specific obligations for high-risk applications common in finance, such as credit scoring and insurance underwriting [10, Art. 6].

For high-risk AI systems, the Act imposes stringent requirements, including:

1. Maintaining detailed documentation of the AI system's development and functioning;

2. Ensuring human oversight throughout the AI system's lifecycle;

3. Conducting conformity assessments before placing systems on the market;

4. Implementing robust risk management processes.

This approach provides a flexible framework that can adapt to evolving AI technologies. However, its broad scope may create compliance challenges for financial institutions, particularly in determining risk classifications for complex AI systems and implementing the required oversight mechanisms.

The AI Act complements existing EU legislation, such as the General Data Protection Regulation (GDPR) [11], which continues to play a crucial role in governing the use of personal data in AI systems. Article 22 of the GDPR grants individuals the right not to be subject to decisions based solely on automated processing, including profiling, which produces legal or similarly significant effects [11, Art. 22]. This provision has significant implications for AI-driven decision-making in finance, potentially requiring human intervention in high-stakes decisions such as loan approvals or insurance underwriting.

The GDPR's principles of data minimization and purpose limitation [11, Art. 5(1)(c)] pose particular challenges for AI applications in finance, which often rely on large, diverse datasets to improve accuracy and predictive power. Financial institutions must carefully balance the data requirements of their AI systems with these foundational data protection principles, potentially limiting the effectiveness of certain AI models.

The concept of a "right to explanation" for AI-driven decisions, while not explicitly stated in the GDPR, has been inferred from its provisions on automated decision-making [15, p. 733]. This interpretation, if widely adopted, could have far-reaching implications for the design and deployment of AI systems in finance. Financial institutions may need to develop explainable AI techniques that can provide meaningful explanations of complex algorithmic decisions to both regulators and consumers.

In contrast to the EU's comprehensive approach, the United States has relied on a patchwork of existing laws and sector-specific regulations to govern AI in finance. This approach provides flexibility but also creates potential regulatory gaps and uncertainties.

U.S. fair lending laws, including the Equal Credit Opportunity Act (ECOA) and the Fair Housing Act, play a crucial role in governing AI-driven credit decisions. These laws prohibit discrimination in lending based on protected characteristics, which extends to the use of AI in credit decisions [14, p. 3]. The application of the disparate impact doctrine to AI systems raises complex legal questions. Under this doctrine, facially neutral practices that have a disproportionate adverse effect on protected groups may violate fair lending laws, even in the absence of discriminatory intent [22, p. 134].

The challenge lies in applying these traditional legal concepts to sophisticated AI models that may inadvertently generate discriminatory outcomes through complex interactions of seemingly neutral factors. For example, an AI system used for credit scoring might identify correlations between credit risk and factors that serve as proxies for protected characteristics, such as zip code or shopping patterns. While these factors may not explicitly consider race or gender, their use could result in discriminatory outcomes that violate fair lending laws.

The 2019 HUD lawsuit against Facebook for discriminatory housing advertisements illustrates the difficulties in applying fair lending concepts to AI-driven platforms [22, p. 136]. In this case, HUD alleged that Facebook's ad targeting algorithm allowed advertisers to exclude protected groups from seeing housing ads, even when the advertisers did not explicitly choose to do so. This case highlights the need for regulators to develop new frameworks for identifying and addressing algorithmic discrimination in financial services.

U.S. financial regulators have issued various guidance documents addressing the use of AI in finance. The Federal Reserve's SR 11-7 guidance on model risk management, while not AI-specific, has been applied to AI models used in banking [4, p. 18]. This guidance emphasizes the importance of model validation, ongoing monitoring, and effective challenge of models. However, its application to advanced AI techniques, such as deep learning models, raises complex compliance questions. For instance, how can financial institutions effectively validate AI models that continuously learn and adapt based on new data?

The Securities and Exchange Commission (SEC) has issued guidance on the application of the Investment Advisers Act to automated investment advisory programs, commonly known as "robo-advisors" [13, p. 26]. This guidance emphasizes the need for clear disclosures about the limitations of automated advice and the importance of maintaining appropriate compliance programs. However, it does not fully address the unique challenges posed by more advanced AI systems that may adapt and evolve over time, potentially changing their investment strategies without human intervention.

The U.S. approach to AI regulation in finance reflects a broader regulatory philosophy that emphasizes principles-based regulation and relies on existing legal frameworks to address new technological challenges. While this approach provides flexibility and allows for innovation, it may also create uncertainty for financial institutions seeking to implement AI systems. The lack of clear, AI-specific regulations may lead to inconsistent interpretations and enforcement actions across different regulatory agencies.

The United Kingdom has adopted a principles-based approach to AI regulation in finance, focusing on existing regulatory frameworks and sector-specific guidance. The Financial Conduct Authority (FCA) has issued guidance on the use of AI in financial services, emphasizing the importance of explainability, governance, and fairness [8, p. 30, 32].

The FCA's approach places significant responsibility on financial institutions to ensure that their AI systems are explainable and subject to appropriate governance. This principles-based approach provides flexibility for innovation but may lack the specificity needed to address complex AI-related issues. The emphasis on explainability aligns with global trends but raises practical challenges given the complexity of many AI systems.

Key principles emphasized by the FCA include:

1. Transparency: Financial institutions should be able to explain how their AI systems make decisions, particularly when those decisions affect consumers.

2. Accountability: Clear lines of responsibility should be established for AI systems, with senior management held accountable for their development and use.

3. Fairness: AI systems should not lead to unfair discrimination or exclusion of particular groups of consumers.

4. Ethics: The use of AI should align with broader ethical principles and societal values.

The UK's approach is notable for its focus on fostering innovation while managing risks. The FCA's regulatory sandbox, which allows firms to test innovative products and services in a controlled environment, has been particularly influential in shaping the development of AI applications in finance [8, p. 40]. This approach allows regulators to gain hands-on experience with new AI technologies and develop more informed regulatory strategies.

However, the principles-based approach also presents challenges. Financial institutions may struggle to interpret and apply broad principles to specific AI applications, particularly in areas where the technology is rapidly evolving. There is also a risk that without more prescriptive rules, different institutions may interpret the principles inconsistently, leading to uneven application of AI governance standards across the financial sector.

Various international organizations have developed principles and guidelines for AI governance that, while not legally binding, influence regulatory approaches and industry practices. The OECD AI Principles, adopted by OECD member countries, provide a framework for responsible AI development [17, p. 12]. These principles emphasize the importance of:

1. Human-centered values and fairness;
2. Transparency and explainability;
3. Robustness, security, and safety;
4. Accountability.

The Financial Stability Board (FSB) has issued reports on the financial stability implications of AI, highlighting potential risks and regulatory considerations [9, p. 2]. These reports emphasize the need for:

1. Robust governance frameworks for AI systems;
2. Effective risk management practices;
3. Appropriate disclosure and transparency measures.

While these soft law instruments lack formal legal force, they play an important role in shaping regulatory approaches and industry best practices. Their principles-based nature allows for flexibility in implementation but may not provide sufficient guidance for addressing specific legal challenges posed by AI in finance.

The diverse regulatory approaches across jurisdictions reflect the complex and evolving nature of AI technologies in finance. These varying approaches create both opportunities and challenges for financial institutions seeking to implement AI systems while ensuring regulatory compliance.

One of the key challenges in regulating AI in finance is addressing the potential for algorithmic bias and discrimination. AI systems may inadvertently generate discriminatory outcomes by identifying proxies for protected characteristics, challenging traditional legal concepts of discrimination [22, p. 134]. The complexity of AI algorithms may make it difficult to identify and explain discriminatory outcomes, complicating enforcement of anti-discrimination laws [14, p. 17].

AI systems trained on historical data may perpetuate or amplify existing biases, raising questions about liability and the standard of care required in AI development [15, p. 5]. Financial institutions must carefully consider the potential for bias in their training data and implement robust processes for detecting and mitigating such biases.

The legal concept of "disparate impact" takes on new dimensions in the context of AI. Courts and regulators must grapple with how to apply this doctrine to complex AI models where discriminatory outcomes may not be easily traceable to specific inputs or decision rules. The ongoing debate over the use of alternative data in credit decisions exemplifies these challenges, as regulators seek to balance the potential for increased financial inclusion with the risks of unintended discrimination [22, p. 138].

Another significant challenge is ensuring transparency and explainability of AI systems. The "black box" nature of many AI algorithms makes it difficult for financial institutions to provide clear explanations of how decisions are made, particularly in complex areas such as credit scoring or algorithmic trading. This lack of transparency raises concerns about accountability and fairness, as well as potential conflicts with regulatory requirements for explainable decision-making.

The European Banking Authority's report on Big Data and Advanced Analytics highlights these challenges, calling for further guidance on explainability requirements for AI in finance [8, p. 30]. As regulators develop more specific guidelines in this area, financial institutions will need to invest in technical solutions and governance frameworks that can meet evolving explainability standards.

Explainable AI (XAI) techniques are emerging as a potential solution to this challenge. These methods aim to provide interpretable explanations of AI decisions without sacrificing the performance benefits of complex algorithms. However, the development and implementation of XAI techniques in finance raise their own legal and regulatory questions. For example, how much explanation is sufficient to meet regulatory requirements? How can the accuracy and reliability of these explanations be verified?

The use of AI in finance also raises important questions about liability and accountability. Determining responsibility for AI-driven errors or harmful decisions is legally complex, particularly when multiple parties are involved in developing and deploying AI systems [24, p. 15]. Traditional concepts of legal liability may need to be reevaluated in light of AI's capacity for autonomous decision-making.

The concept of "meaningful human control" is emerging in legal discourse as a potential framework for attributing responsibility in AI systems. This concept suggests that humans should maintain a certain level of control and oversight over AI decision-making processes, particularly in high-stakes financial decisions. However, the practical implementation of meaningful human control in complex AI systems remains a challenge.

Data protection and privacy considerations also play a crucial role in the regulation of AI in finance. The data-intensive nature of many AI applications conflicts with principles of data minimization and purpose limitation enshrined in laws like the GDPR [11, Art. 5(1)(c)]. Financial institutions must carefully balance the data requirements of their AI systems with legal obligations to collect only the data necessary for specified purposes.

The ability of AI systems to derive unexpected insights from data also challenges traditional notions of consent and purpose limitation. For example, an AI system designed for fraud detection might identify patterns in customer behavior that are also indicative of certain health conditions. The use of such insights, while potentially valuable for risk assessment, raises complex legal and ethical questions about the permissible use of data beyond its original purpose.

As AI systems become more sophisticated and interconnected, there is also growing concern about their potential to introduce new forms of systemic risk to the financial system. The use of AI in high-frequency trading, for instance, could potentially lead to unexpected market volatility or flash crashes if multiple AI systems interact in unforeseen ways. Regulators will need to develop new approaches to monitor and mitigate these systemic risks, potentially including stress testing of AI systems and enhanced coordination between financial institutions and regulators.

The global nature of financial markets and AI development also raises challenges for regulatory harmonization. While international standards and soft law instruments provide some common ground, the divergent regulatory approaches across jurisdictions create potential for regulatory arbitrage and compliance challenges for multinational financial institutions. Efforts to promote international cooperation and regulatory convergence in AI governance will be crucial to addressing these challenges.

Another important aspect of AI regulation in finance is the need for ongoing monitoring and adaptation of regulatory frameworks. As AI technologies continue to evolve rapidly, regulators must develop mechanisms to stay informed about emerging risks and opportunities. This may involve closer collaboration with industry stakeholders, academic researchers, and technology experts to ensure that regulatory approaches remain relevant and effective.

The role of human oversight in AI-driven financial systems is also a critical area of regulatory focus. While AI can enhance decision-making processes, there is growing recognition of the importance of maintaining human judgment and accountability, particularly in high-stakes financial decisions. Regulators are grappling with how to define appropriate levels of human oversight and intervention in AI systems, balancing the benefits of automation with the need for human expertise and ethical considerations [6, p. 15].

The use of AI in regulatory compliance itself, often referred to as "RegTech," presents both opportunities and challenges. AI-powered compliance tools can help financial institutions more efficiently monitor transactions, detect potential violations, and generate regulatory reports. However, the use of AI in compliance also raises questions about the reliability and auditability of these systems, as well as potential conflicts of interest if the same AI technologies are used for both business operations and regulatory compliance [16, p. 3].

The emergence of decentralized finance (DeFi) and blockchain-based financial systems powered by AI introduces new regulatory challenges. These systems often operate outside traditional financial infrastructures, making it difficult for regulators to apply existing frameworks. The intersection of AI, blockchain, and finance raises complex questions about jurisdiction, accountability, and the application of traditional financial regulations to decentralized systems [23, p. 25].

Moreover, the use of AI in cross-border financial transactions and international financial markets presents unique regulatory challenges. Different jurisdictions may have conflicting requirements for data protection, algorithmic transparency, and AI governance. Financial institutions operating globally must navigate this complex landscape, potentially leading to increased compliance costs and operational complexities [2, p. 7].

The role of AI in financial stability and systemic risk management is another area of growing regulatory concern. While AI can enhance risk assessment and early warning systems, it may also introduce new forms of systemic risk. For instance, if multiple financial institutions rely on similar AI models for risk assessment or trading strategies, it could lead to herding behavior and amplify market volatility [9, p. 3]. Regulators are exploring ways to assess and mitigate these AI-driven systemic risks, including stress testing AI systems and developing new supervisory approaches for AI-driven financial activities.

The ethical implications of AI in finance extend beyond issues of fairness and non-discrimination. For example, the use of AI in investment management raises questions about the fiduciary duty of financial advisors and the extent to which AI systems can be trusted to act in clients' best interests. Regulators are grappling with how to apply traditional concepts of fiduciary responsibility to AI-driven investment advice [13, p. 28].

Furthermore, the potential for AI to exacerbate financial inequality is a growing concern. While AI has the potential to expand access to financial services for underserved populations, it may also widen the gap between those who have access to sophisticated AI-driven financial tools and those who do not. Regulators are considering how to ensure that the benefits of AI in finance are distributed equitably and do not lead to further financial exclusion [22, p. 140].

The use of AI in credit scoring and lending decisions has drawn particular regulatory scrutiny. Traditional credit scoring models have long been criticized for perpetuating historical biases and excluding certain groups from access to credit. AI-powered credit scoring systems promise to use a wider range of data points to make more accurate and inclusive lending decisions. However, these systems also raise concerns about privacy, fairness, and the potential for new forms of discrimination [15, p. 735].

Some jurisdictions are exploring regulatory sandboxes specifically focused on AI in finance.

These controlled environments allow financial institutions to test innovative AI applications under regulatory supervision, providing valuable insights for both firms and regulators. The success of these initiatives could inform future regulatory approaches and help bridge the gap between innovation and regulation [8, p. 40].

The cybersecurity implications of AI in finance are also a key area of regulatory focus. AI systems can be vulnerable to adversarial attacks, where malicious actors manipulate input data to deceive the AI model. In the context of financial services, such attacks could have severe consequences, potentially leading to market manipulation or financial fraud. Regulators are exploring how to incorporate AI-specific cybersecurity considerations into existing financial sector cybersecurity frameworks [20, p. 10].

The governance of AI models in finance is another critical regulatory challenge. As AI models become more complex and autonomous, traditional governance structures may not be sufficient. Regulators are considering new approaches to AI governance, including requirements for AI ethics committees, regular algorithmic audits, and enhanced board-level oversight of AI strategies [4, p. 22].

The use of AI in financial crime prevention, while promising, also raises regulatory concerns. AI can significantly enhance anti-money laundering (AML) and know-your-customer (KYC) processes, but it also introduces new risks. For example, sophisticated criminals might use AI to develop more advanced money laundering techniques that can evade AI-powered detection systems. Regulators are working to ensure that AI-driven financial crime prevention measures are robust, adaptive, and subject to appropriate oversight [3, p. 5].

The potential for AI to disrupt traditional financial intermediaries and market structures is another area of regulatory interest. As AI enables more direct and personalized financial services, it may challenge the role of traditional banks and financial institutions. Regulators are considering how to adapt existing regulatory frameworks to ensure financial stability and consumer protection in an AI-driven financial landscape [7, p. 10].

The use of natural language processing (NLP) and AI in financial reporting and disclosure is emerging as a new frontier in financial regulation. AI can potentially enhance the accuracy and timeliness of financial disclosures, but it also raises questions about the reliability of AI-generated reports and the potential for manipulation. Regulators are exploring how to ensure the integrity of AI-driven financial reporting while leveraging its potential benefits [18, p. 15].

As AI becomes more prevalent in financial services, there is growing recognition of the need for AI literacy among financial regulators and supervisors. Many jurisdictions are investing in training programs and technical resources to ensure that regulatory staff have the necessary skills to effectively oversee AI-driven financial systems [5, p. 7].

The environmental impact of AI in finance is an emerging area of regulatory concern. The energy consumption of large AI models and data centers used in financial services can be significant. As environmental, social, and governance (ESG) factors become more important in financial regulation, the carbon footprint of AI systems may come under increased scrutiny [21, p. 137].

International coordination in AI regulation for finance remains a significant challenge. While bodies like the Financial Stability Board are working to promote regulatory harmonization, significant differences remain between jurisdictions. These differences could lead to regulatory arbitrage, where financial institutions choose to operate in jurisdictions with more lenient AI regulations [19, p. 8].

The role of AI in central bank digital currencies (CBDCs) is another area of emerging regulatory interest. As many central banks explore the potential of CBDCs, the use of AI in managing and monitoring these new forms of digital money raises novel regulatory questions about privacy, monetary policy, and financial stability [3, p. 7].

**Conclusion**

The regulation of AI in financial services presents a complex and evolving landscape of legal and compliance challenges. The diverse regulatory approaches adopted by different jurisdictions reflect the difficulty of balancing innovation with consumer protection and market stability. As AI technologies continue to advance and reshape the financial industry, regulators and financial institutions will need to collaborate closely to develop effective governance frameworks that can adapt to emerging risks and opportunities.

The key challenges in regulating AI in finance include addressing algorithmic bias and discrimination, ensuring transparency and explainability of AI systems, managing data protection and privacy concerns, mitigating new forms of systemic risk, and promoting international regulatory harmonization. These challenges are compounded by the rapid pace of technological change and the global nature of financial markets.

Moving forward, regulators will need to adopt more agile and adaptive approaches to keep pace with AI innovation in finance. This may involve greater use of regulatory sandboxes, closer collaboration with industry and academia, and investment in AI literacy among regulatory staff. Financial institutions, for their part, will need to prioritize ethical AI development, robust governance frameworks, and proactive engagement with regulators to navigate this complex landscape.

As AI continues to transform financial services, it is clear that traditional regulatory paradigms will need to evolve. The future of AI regulation in finance will likely involve a combination of principle-based approaches to provide flexibility, and specific rules to address high-risk applications. Importantly, regulators must strive to create frameworks that not only mitigate risks but also harness the potential of AI to improve financial inclusion, enhance market efficiency, and contribute to overall financial stability.

Ultimately, the successful regulation of AI in financial services will require a delicate balance between fostering innovation and protecting the public interest. It will demand ongoing dialogue between regulators, financial institutions, technology providers, and other stakeholders to ensure that AI is deployed responsibly and ethically in the financial sector. As we navigate this new frontier, continued research, international cooperation, and adaptive policymaking will be crucial in shaping a financial future where AI serves as a force for positive transformation.

Mirişli Şahmar Sakit oğlu

## MALİYYƏ XİDMƏTLƏRİNDƏ SÜNİ İNTELLEKTİN TƏNZİMLƏNMƏSİ: HÜQUQİ ÇƏRÇİVƏLƏR VƏ UYĞUNLUQ PROBLEMLƏRİNİN TƏHLİLİ

### XÜLASƏ

Bu məqalədə maliyyə xidmətlərində süni intellektin (Sİ) tənzimlənməsi üçün mövcud hüquqi çərçivələr hərtərəfli təhlil edilir və bu texnologiyaların yaratdığı unikal riskləri həll etmək qabiliyyəti qiymətləndirilir. Mövcud ədəbiyyatı, tənzimləyici yanaşmaları və praktiki nümunələri geniş şəkildə araşdıraraq, məlumatların qorunması, alqoritmik ədalət, şəffaflıq və hesabatlılıq kimi əsas problemlər müəyyən və təhlil edilir. Tədqiqat mövcud tənzimləyici çərçivələrdə əhəmiyyətli boşluqlar aşkar etmişdir və istehlakçıların qorunmasını və bazar sabitliyini təmin edərkən innovasiyanı təşviq edən adaptiv, texnologiyadan asılı olmayan qaydaların təcili ehtiyacını ortaya qoymuşdur. Müxtəlif yurisdiksiyalardakı yanaşmaları müqayisə edərək, yüksək riskli Sİ tətbiqləri üçün xüsusi təlimatlı prinsipə əsaslanan nəzarətin birləşməsi Sİ-nin potensial risklərini azaltmaqla yanaşı innovasiyanı təşviqetmək üçün vacib olduğu müəyyən edilmişdir. Bu məqalə maliyyə sahəsində Sİ-nin idarə edilməsi üzrə davam edən müzakirələrə töhfə verir və siyasətçilər və maliyyə qurumları üçün texnologiya, maliyyə və hüququn mürəkkəb kəsişməsində yol göstərən incə bir çərçivə təklif edir.

Миришли Шахмар Сакит оглу

## РЕГУЛИРОВАНИЕ ИСКУССТВЕННОГО ИНТЕЛЛЕКТА В ФИНАНСОВЫХ УСЛУГАХ: АНАЛИЗ ПРАВОВЫХ РАМОК И ПРОБЛЕМ СООТВЕТСТВИЯ

### РЕЗЮМЕ

В данной статье представлен всесторонний анализ существующих правовых рамок регулирования искусственного интеллекта (ИИ) в сфере финансовых услуг и оценивается их эффективность в решении уникальных рисков, создаваемых этими технологиями. Посредством обширного обзора существующей литературы, регуляторных подходов и практических примеров мы выявляем и оцениваем ключевые области озабоченности, включая защиту данных, алгоритмическую справедливость, прозрачность и подотчетность. Исследование выявляет значительные пробелы в существующих нормативных рамках и подчеркивает острую необходимость в адаптивных, технологически нейтральных правилах, которые могут эффективно решать уникальные риски, создаваемые ИИ, одновременно способствуя инновациям и обеспечивая защиту потребителей и стабильность рынка. Сравнивая подходы в различных юрисдикциях, мы утверждаем, что сочетание принципиального надзора с конкретными руководствами для высокорисковых приложений ИИ имеет важное значение для стимулирования инноваций при одновременном смягчении потенциальных рисков ИИ. Эта статья вносит вклад в продолжающийся диалог по управлению ИИ в финансах, предлагая нюансированную структуру для политиков и финансовых учреждений для навигации в сложном пересечении технологий, финансов и права.